# Enhancing the Capture of Magnetic Nanoparticles Inside of Ferromagnetic Nanostructures Using External Magnetic Fields

Reyne Dowling[1] and Mikhail Kostylev[1]

[1]School of Physics, University of Western Australia, Perth, WA, Australia, reyne.dowling@research.uwa.edu.au

**The influence of an external magnetic field upon the capture of 130 nm magnetic nanoparticles by ferromagnetic nanostructures was investigated through simulations and experiments. The magnetophoretic forces acting upon a nanoparticle were simulated in external magnetic fields parallel and perpendicular to a ferromagnetic nanostructure consisting of an array of antidots or dots. Changing the direction of the external field was found to dramatically alter the magnetophoretic forces acting on the particle and the trajectories of the MNPs. A field parallel to the nanostructures' surfaces generated magnetophoretic forces that directed the nanoparticle into the nanostructures. A perpendicular field produced forces directing the particle onto the surface of the structures. Three sets of nanostructures were etched into the surfaces of Permalloy films using ion beam lithography. Magnetic nanoparticles were then deposited onto the surfaces of the films under a parallel or perpendicular external magnetic field. The number and distribution of particles in the nanostructures was then analysed to obtain the capture efficiencies of each structure which indicate the proportion of particles trapped inside. Without an external field, the highest efficiency of 70.8% was displayed by arrays of circular antidots with array of circular dots displaying the lowest of 21.3%. Generally, arrays of antidots displayed higher capture efficiencies than arrays of dots. Addition of an external field parallel to the surface significantly increased the capture efficiencies with the lowest now 44.1% for arrays of circular dots. Conversely, addition of a field perpendicular to the surface decreased the efficiencies with the lowest at 7.5% for arrays of circular dots. Under this field, the particles were instead caught on the outer edges of the nanostructures. These results suggest that application of a parallel external magnetic field promotes the capture of magnetic nanoparticles within ferromagnetic nanostructures. Application of a perpendicular field increases the capture of nanoparticles onto the outer surface and edges of nanostructures.**

*Index Terms*—Ferromagnetic materials, Superparamagnetic iron oxide nanoparticles

## I. Introduction

Magnetic nanoparticles (MNPs) are crucial tools in emerging and rapidly growing nanotechnologies due to the versatility of their coatings and the ability to manipulate them using magnetic fields [1-3]. Of particular interest are their applications in magnetic particle assays in the emerging micro total analytical systems (μTAS) used in medical diagnostics and food sampling [4-8]. These systems are capable of analysing samples faster, cheaper and easier than traditional assay methods such as ELISA and fluorescence imaging [9]. These biosensors are also compact and portable allowing for testing in homes and remote locations. These advantages make μTAS suitable for detection of cancer and infectious diseases such as SARS-CoV-2 during pandemics [10-13]. In most μTAS, the nanoparticles are functionalised with antibody or protein coatings that bind to specific analytes of interest such as viruses and cancer biomarkers. An external magnetic field is then used to direct the bound magnetic nanoparticles onto sensing surfaces on which magnetic assays are performed using techniques such as giant magnetoresistance (GMR) or nuclear magnetic resonance (NMR) [12, 14-16]. Generally, the sensitivity of an assay will depend upon the number of particles in the sensing area. It is therefore useful to know where nanoparticles are going to be caught within the complex nanostructures of a device. A diverse variety of nanostructures have been used to capture MNPs onto the sensing area such as nanowires, membranes and crystal scaffolds [17-19]. Sushruth et al. trapped magnetic nanoparticles within arrays of antidots etched into a ferromagnetic layer [20, 21]. The particles were then detected using ferromagnetic resonances in the ferromagnetic nanostructure. This study extends this work by comparing the capture of MNPs in different ferromagnetic array nanostructures. The results will also shed a light on interactions between nanoparticles and ferromagnetic nanostructures in general. Choosing a nanostructure that optimises the capturing of magnetic nanoparticles is of particular interest for improving the efficiency and sensitivity of magnetic particle assays.

Another factor that may influence the ability of a ferromagnetic nanostructure to capture MNPs is the application and orientation of any external magnetic fields. Non-uniform magnetic fields result in magnetophoresis, a process in which the magnetic nanoparticles are drawn along the gradients of the local magnetic field produced by the nanostructure [22, 23]. Many biosensors already exploit the magnetophoretic forces generated by non-uniform, external magnetic fields produced by permanent magnets to guide nanoparticles onto sensing nanostructures and improve the capture efficiency [11, 24-26]. Ezzaier et al. found that an external magnetic field had a stronger influence on the capture of magnetic nanoparticles than the geometry of a micropillar nanostructure [27]. Magnetic nanoparticles will experience magnetophoresis near a ferromagnetic nanostructure as the structure will possess an intrinsically non-uniform local magnetic field since the magnetisation is discontinuous [28]. Little et al. attempted to use these forces to capture MNPs onto the edges of a GMR sensor with mild success [29]. Without an external magnetic field, the local magnetic field of a ferromagnetic nanostructure is weak due to the net magnetisation vanishing in the bulk of the material, due to the presence of a domain structure. Ferromagnetic structures therefore produce negligible magnetophoretic forces without an external field. Application of an external magnetic field aligns the magnetic moments, producing a stronger local magnetic field around the ferromagnetic nanostructure. By applying an external magnetic field, one can modify the magnetophoretic forces acting upon magnetic nanoparticles to influence where each particle is captured. Note that the Ferromagnetic Resonance method of MNP detection requires application of a significant external field in order to magnetise the sensing antidot structure to saturation [20, 21]. As follows from the above, this may be beneficial for MNP capturing and potentially allows for *simultaneous* capturing and detection of magnetic nanoparticles. This is difficult to achieve when employing a GMR detection method, for which detection is carried out in a small or vanishing field.

In this study, the influences of the nanostructure geometry and external magnetic field on MNP capturing were investigated by observing the number and location of MNPs caught within four different nanostructure geometries. These geometries consisted of arrays of circular and square antidots and dots etched into a thin Permalloy film. Distributions of nanoparticles in all four geometries were also obtained after capturing under uniform, external magnetic fields oriented parallel and perpendicular to the film surface. The distributions were compared to determine whether application of an external magnetic field could be used to improve the capturing efficiency. The magnetophoretic forces acting on the MNPs and the resulting MNP trajectories were simulated for each geometry based upon micromagnetic simulations of the magnetic state of the nanostructures.

## II. Materials and Methods

MNP distributions were first obtained for nanoparticles captured inside ferromagnetic nanostructures in the absence of an external magnetic field. To fabricate the ferromagnetic nanostructures, a 30 nm-thick layer of Permalloy was coated onto 5 mm square pieces of silicon wafer using magnetron sputtering in an Argon atmosphere. Magnetron sputtering was performed on the silicon substrates at a temperature of 150 °C and Argon pressure of 6 mTorr. The base pressure in the chamber was lower than $10^{-7}$ Torr.

Four nanostructure geometries were chosen for this study, consisting of 15 × 15 square arrays of 400 nm square dots and 400 nm diameter circle dots and their inverses (circle and square antidots). The elements in the array were separated by 200 nm in the x and y directions. Twenty nanostructures of each geometry were then etched into all three films via focused ion beam (FIB) lithography using the FEI Helios FIB-SEM at the Centre for Microscopy, Characterisation and Analysis (CMCA) at the University of Western Australia [30-33]. With an accelerating voltage of 30 kV and an ion current of 80 pA, the ion beam spot size was approximately 21 nm. This beam was capable of cleanly etching the desired shapes at the desired 30 nm depth. Each nanostructure was located approximately 9 μm apart to reduce the influence of the magnetic fields of neighbouring structures. Once etching was completed each nanostructure was imaged using a FEI Verios XHR scanning electron microscope (SEM) at the Centre for Microscopy and Characterisation Analysis (CMCA).

The magnetic nanoparticles used were 130 nm-diameter iron-oxide nanoparticle clusters (Nanomag®-D, Micromod Partikeltechnologie GmbH) coated in dextran. The MNPs were purchased in an aqueous colloidal suspension with a concentration of 25 mg/mL, which was diluted to 0.01 mg/mL using deionised (DI) water. A single 4 μL droplet of diluted nanoparticle solution was deposited onto the surface of each film using a micropipette. The nanostructures were located directly below the centre of the droplets. The MNPs suspended in the droplets are captured by the nanostructures as the droplet evaporates. After deposition, the nanostructures and captured MNPs were imaged using the SEM. The MNPs caught by each nanostructure were counted and sorted according to whether they were inside or outside of the antidot inclusions, as indicated in Figure 1. However, the dot geometries have significantly larger areas compared to the antidot geometries in which more MNPs could be caught. To account for the different volumes of material removed from the Permalloy films for each geometry, the average numbers of particles inside and outside the nanostructure, $N_{IN}$ and $N_{OUT}$, were scaled by the values $S_{IN}$ and $S_{OUT}$ respectively, where

$$S_{IN} = \frac{volume\ remaining}{volume\ removed} \qquad \text{Eq. 1}$$

And

$$S_{OUT} = \frac{volume\ removed}{volume\ remaining}. \qquad \text{Eq. 2}$$

Finally, the capture efficiency of the arrays was calculated using the definition below. This number is indicative of the nanostructure's ability to capture MNPs internally and was compared amongst the four geometries. The capture efficiency can also be used to estimate the proportion of MNPs that would be captured inside or outside of each nanostructure geometry.

$$Capture\ Efficiency = \frac{S_{IN}\ N_{IN}}{S_{IN}\ N_{IN} + S_{OUT}\ N_{OUT}} \qquad \text{Eq. 3}$$

To investigate the effect of an external magnetic field on the nanoparticle capturing this process was repeated for two new films with deposition now occurring within external magnetic fields oriented parallel to and perpendicular to the film. A magnetic field with perpendicular orientation was achieved by placing a 1.5 cm square permanent magnet of width 4 mm roughly 2 mm below the nanostructures. The film was placed onto the magnet with the nanostructures located above the centre of the magnet. At this position, the magnetic field produced by the magnet was approximately 3.5 kOe. For a parallel external field, the film was placed between two parallel cylindrical permanent magnets with the nanostructures being positioned equidistant from both magnets and along the central axis of the magnets, as shown in Figure 2. The strength of the field at the nanostructures' positions was 1.38 kOe.

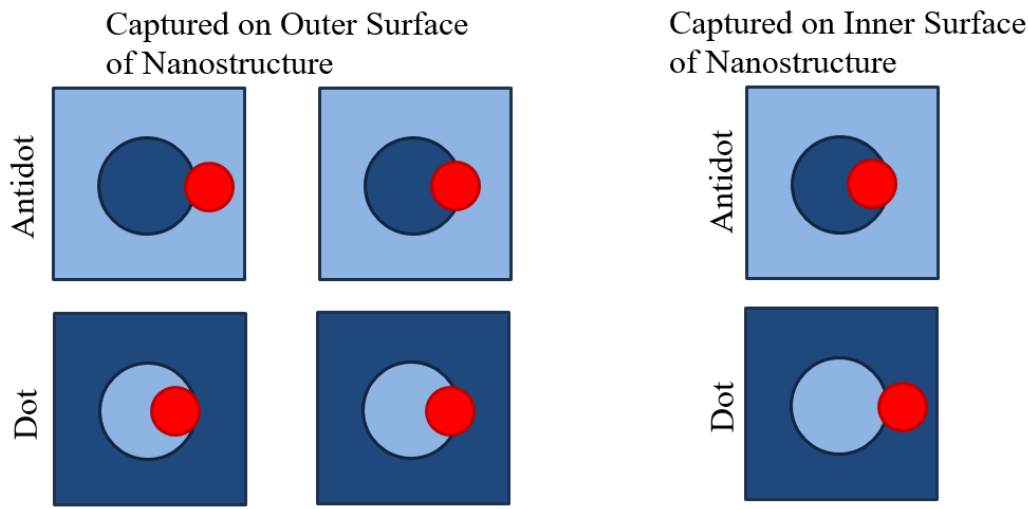

Figure 1: Diagrams depicting the position of a magnetic nanoparticle (red) with respect to an antidot or dot and indicating whether these nanoparticles are considered to have been captured on the outer or inner surface of the nanostructure. Lighter blue regions represent areas containing Permalloy while darker regions represent empty spaces in which Permalloy has been removed from the film. For a nanoparticle to be considered as caught inside of the nanostructure, most of the nanoparticle must be located in one of the darker regions of the nanostructure. Otherwise, the particle is counted as captured on the outside of the nanostructure.

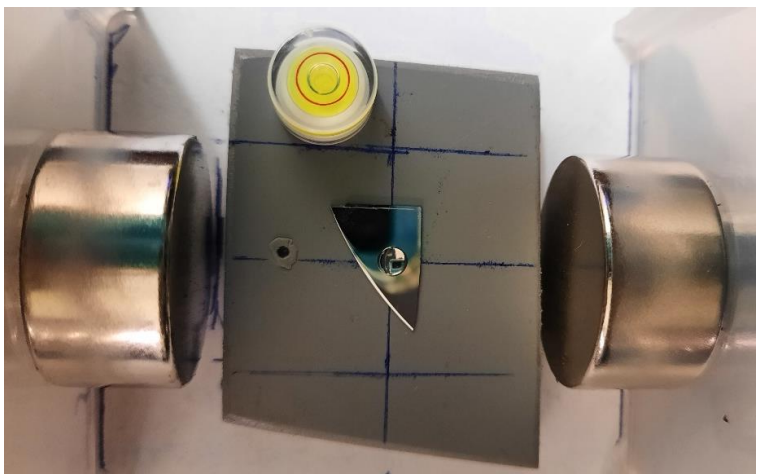

Figure 2: A droplet of suspended magnetic nanoparticles after being deposited onto the surface of a Permalloy thin film via a pipette. The droplet is positioned directly over antidot and dot nanostructures hollowed into the surface of the Permalloy film and equidistant between two cylindrical neodymium permanent magnets. The magnetic field at the droplet had a strength of approximately 1.38 kOe and was aligned in the plane of the film, along the y-axis.

III. THEORY AND CALCULATIONS

The magnetisations for the four nanostructures with external field oriented parallel and perpendicular was modelled using the micromagnetic simulation software MuMax$^3$ [34, 35]. Figure 3 illustrates the geometry and dimensions of the simulated array of circular antidots. These geometries consisted of $3 \times 3$ arrays rather than $5 \times 5$ arrays to reduce the computation time. The results for the reduced arrays are expected to be qualitatively similar to the results for arrays of any larger size. The array nanostructures were divided into 3D meshes with each $5 \times 5 \times 10$ nm cell containing a single magnetic moment. In these simulations, the nanostructure surface is parallel to the x - y plane with the z axis indicating distance from the surface.

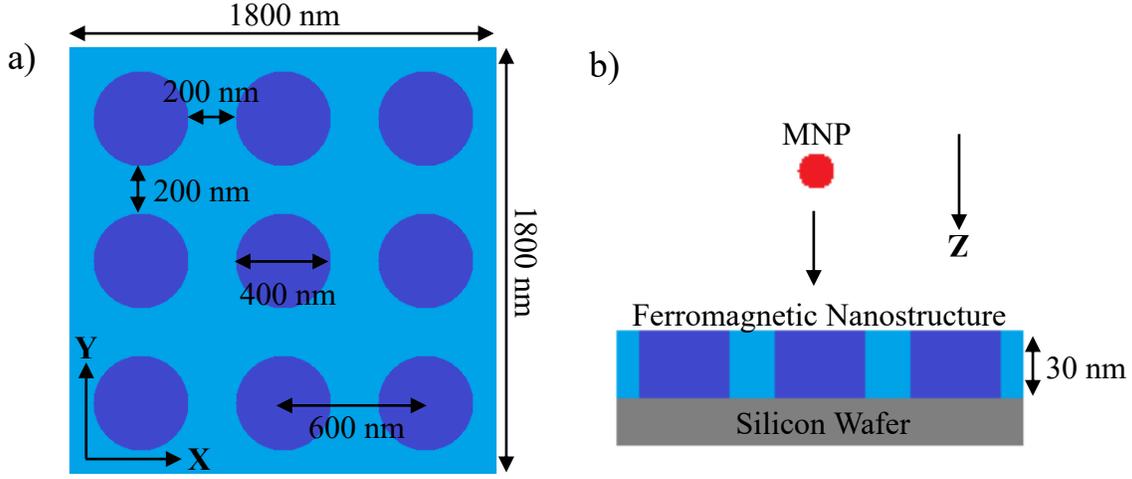

*Figure 3: a) The geometry and dimensions of the simulated 3 x 3 circular antidot array nanostructure. Darker areas indicate empty spaces where material was removed from the film (the inverse is true for dot arrays). b) A cross-section of the 30 nm thick nanostructure with a magnetic nanoparticle located above the surface of the film. The MNP is attracted towards the surface of the nanostructure by magnetophoretic forces.*

Magnetisations were obtained for nanostructures located within external magnetic fields of 2 kG magnitude directed parallel (along the y axis) to and perpendicular (along the z axis) to the film surface. To calculate what local magnetic field will be generated from these simulated magnetisations, one can consider each magnetic moment to be a point dipole producing the magnetic field

$$\overrightarrow{B_{dipole}} = \frac{3\vec{r}(\vec{m}\cdot\vec{r})}{r^5} - \frac{\vec{m}}{r^3} \qquad \text{Eq.4}$$

Where $\vec{r}$ is the vector from the dipole to the location at which the magnetic field is calculated and $\vec{m}$ is the magnetisation vector for the point dipole. The magnetic fields of each moment must be summed together to get the total magnetic field at any point due to the magnetic moments:

$$\vec{B} = \overrightarrow{B_{ex}} + \sum \overrightarrow{B_{dipole}} \qquad \text{Eq.5}$$

In the absence of an external magnetic field $\overrightarrow{B_{ex}}$, the net magnetisation is zero in the bulk of the ferromagnet as the magnetic structure is separated into domains. Only weak local magnetic fields, known as demagnetising fields, are produced by the moments near the edges of the nanostructure where the magnetisation is discontinuous. When exposed to an external magnetic field, the net magnetisation is non-zero as moments align with the external field, breaking the domain structure. The demagnetising field of the structure is far stronger and the bulk of the structure now contributes to the local magnetic field. Simulated z-components of the magnetic fields for an array of circular antidots in parallel and perpendicular external magnetic fields are shown in Figures 4 and 5 respectively. The local fields produced under these external magnetic fields are very different which demonstrates that the local magnetic field is strongly dependent upon the geometry of the ferromagnet and the direction of the external magnetic field. Note that a 2 kG magnetic field applied in the direction perpendicular to the nanostructure plane is not enough to saturate the ferromagnetic nanostructure. This is because the saturation magnetisation of Permalloy, $4\pi M_s$, is known to be significantly larger - around 10.5 kG. The demagnetising field that arises due to film nanostructuring does not reduce the field needed to saturate the nanostructure very significantly [36]. Therefore, canting of the magnetisation vectors in the perpendicular-to-plane direction is expected when the 2 kG perpendicular magnetic field is applied.

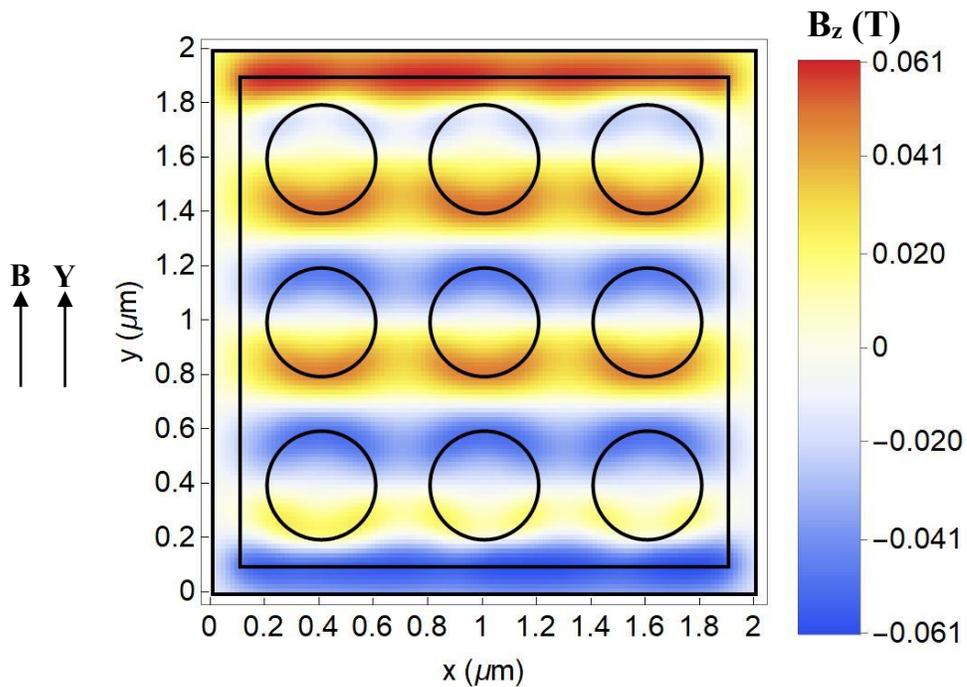

*Figure 4: Two-dimensional plot of the z-component of the total magnetic field produced by a ferromagnetic nanostructure measured in Tesla. The magnetic field has been calculated from a distance of 90 nm above the surface of the nanostructure. Circles have also been plotted to indicate where the antidots are located on the nanostructure surface. The external magnetic field B is directed along the y axis with a magnitude of 2 kG.*

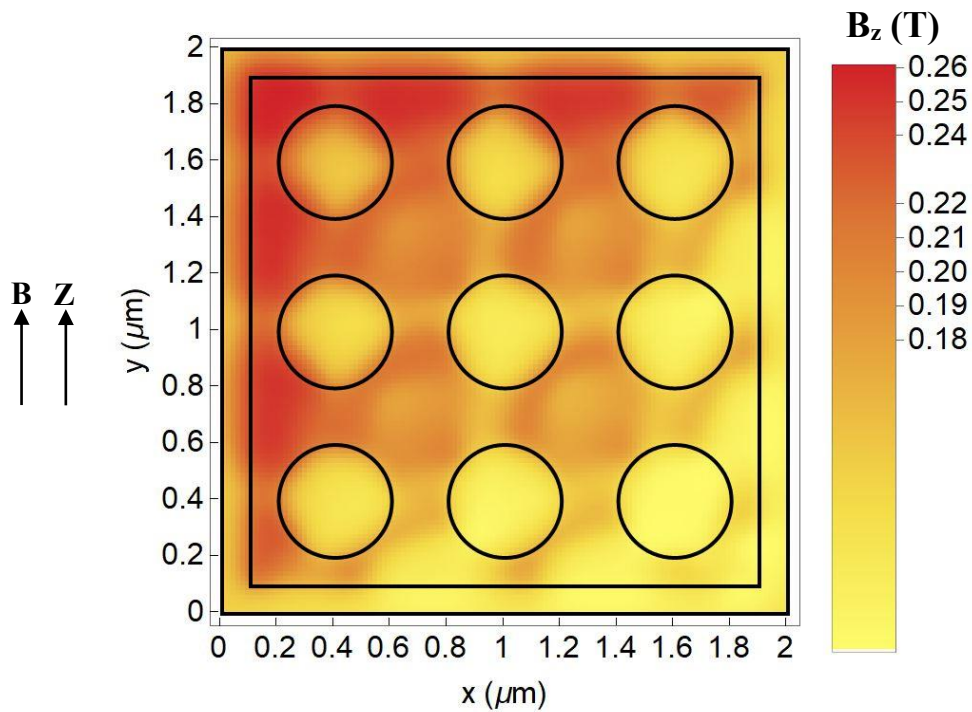

*Figure 5: Two-dimensional plot of the z-component of the total magnetic field produced by a ferromagnetic nanostructure measured in Tesla. The magnetic field has been calculated from a distance of 90 nm above the surface of the nanostructure. Circles have also been plotted to indicate where the antidots are located on the nanostructure surface. The external magnetic field B is directed along the z axis with a magnitude of 2 kG.*

Magnetophoresis is a process in which magnetic forces guide magnetic nanoparticles along the gradient lines of a magnetic field. The magnetophoretic force acting on each superparamagnetic nanoparticle can be calculated [37] using the magnetic field from Eq. 5 using the equation

$$\vec{F_m} = A\,\vec{\nabla}|\vec{B}|^2 = A\,\vec{\nabla}(B_x^2 + B_y^2 + B_z^2) = 2\,A\,B_x\vec{\nabla}B_x + 2\,A\,B_y\vec{\nabla}B_y + 2\,A\,B_z\vec{\nabla}B_z \qquad \text{Eq.6}$$

Where

$$A = \frac{V_{mnp}\,(\chi_{mnp} - \chi_{fluid})}{2\,\mu_0} \qquad \text{Eq.7}$$

Where $V_{mnp}$ is the volume of the magnetic nanoparticle, $\mu_0$ is the vacuum permeability and $\chi_{mnp}$ and $\chi_{fluid}$ are the magnetic susceptibilities of the particle and fluid respectively. For these calculations, the fluid was chosen to be water with a susceptibility of $9.04 \times 10^{-6}\,m^3$. A complete derivation of each magnetophoretic force component can be found in Appendix B. This equation assumes that due to being superparamagnetic, each nanoparticle can be considered to be a single domain behaving as if it were a magnetic moment located at a point. To account for the size of the nanoparticles, the average force was calculated for points along the diameter of the nanoparticle parallel with the external magnetic field. However, calculating the magnetophoretic force acting on the entire volume of the particle and allowing for rotation of the particles would be computationally expensive and as such has been neglected in these simulations. In addition, only the results for the nanostructure with an array of circular antidots will be shown. The simulated results for the other geometries are qualitatively similar. The calculated z-component of the magnetophoretic force, which attracts or repels magnetic nanoparticles from the nanostructure's surface, is displayed in Figures 6 - 10 for both external magnetic field configurations. The x and y components of the magnetophoretic forces in both configurations can be found in Supplementary Marterial C. Directing the external magnetic field parallel to the surface breaks the symmetry of the system, resulting in a magnetophoretic force that is asymmetric about the x and y axes. However, the perpendicular magnetic field does not break symmetry and the resulting force is symmetric. In a parallel external magnetic field, the magnetophoretic force is attractive above the antidots as shown in Figures 6, 7, and 8. Between the antidots the magnetophoretic force can be attractive or repulsive. Under these magnetophoretic forces, most of the MNPs would be directed towards the antidots.

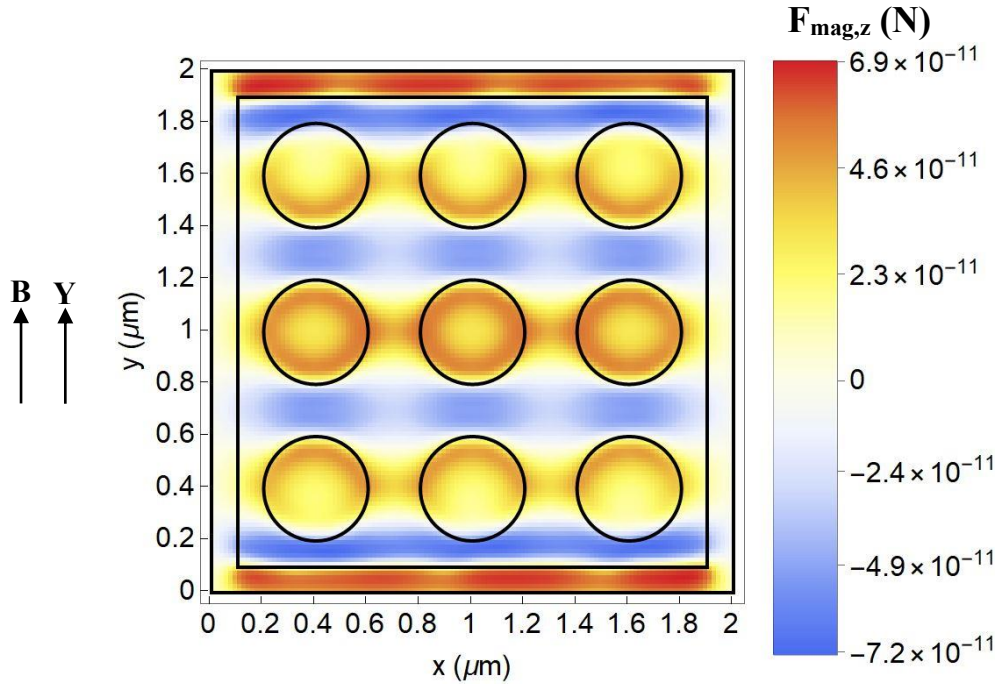

*Figure 6: Two-dimensional plot of the z-component of the magnetophoretic force acting on a 150 nm magnetic nanoparticle, measured in Newtons. The nanoparticle is located 90 nm above the surface of the nanostructure. Circles have also been plotted to indicate where the antidots are located on the nanostructure surface. The external magnetic field B is directed along the y axis with a magnitude of 2 kG.*

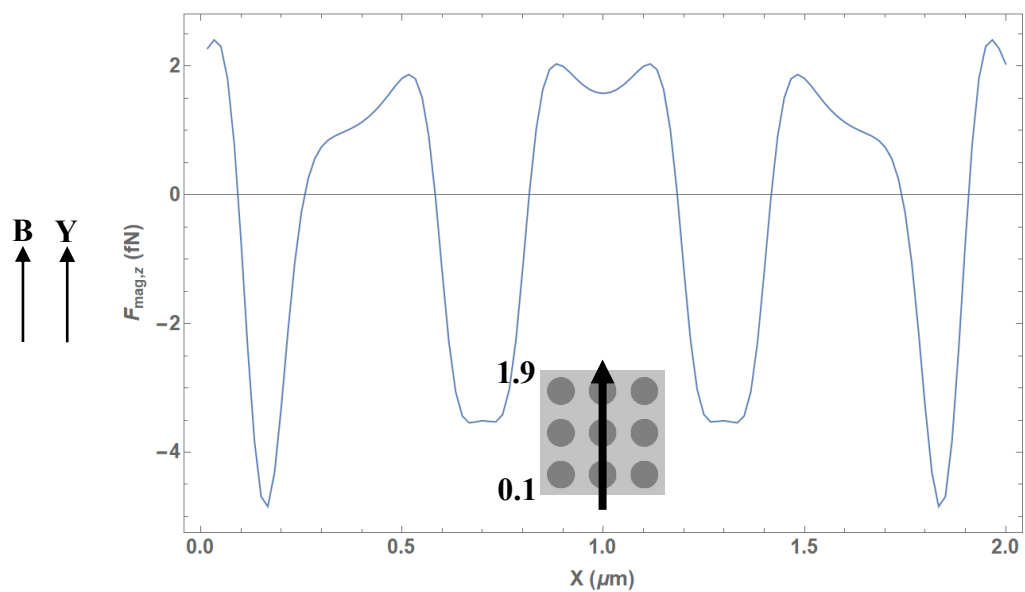

*Figure 7: Simulated z-component of the magnetophoretic force for a 150 nm magnetic nanoparticle located anywhere along the vertical bisection of the nanostructure shown in the inset. The nanoparticle is located 300 nm above the surface of the nanostructure located at z = 0. The external magnetic field B is directed along the y axis with a magnitude of 2 kG. Insert to the figure shows the cross-section along which the force profile was calculated. Negative values of the force correspond to MNP repulsion by the magnetic nanostructure while positive values indicate attraction.*

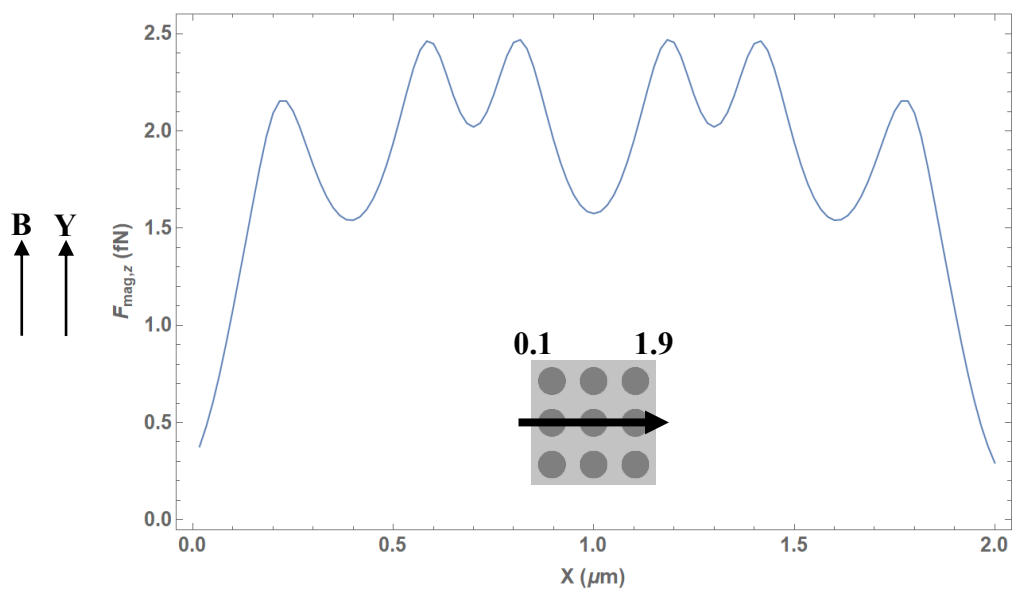

*Figure 8: Simulated z-component of the magnetophoretic force for a 150 nm magnetic nanoparticle located anywhere along the horizontal bisection of the nanostructure shown in the inset. The nanoparticle is located 300 nm above the surface of the nanostructure located at z = 0. The external magnetic field B is directed along the y axis with a magnitude of 2 kG. Insert to the figure shows the cross-section along which the force profile was calculated. Positive values of the force correspond to MNP attraction to the magnetic nanostructure.*

The results presented in Figures 9 and 10 show that under a perpendicular external field, the modelled force is repulsive above the antidots and attractive elsewhere. Such forces would direct most of the nanoparticles towards the outside surface of the nanostructure and away from the antidots. In this configuration, the magnetophoretic force is strongest at the outermost edges of the nanostructure.

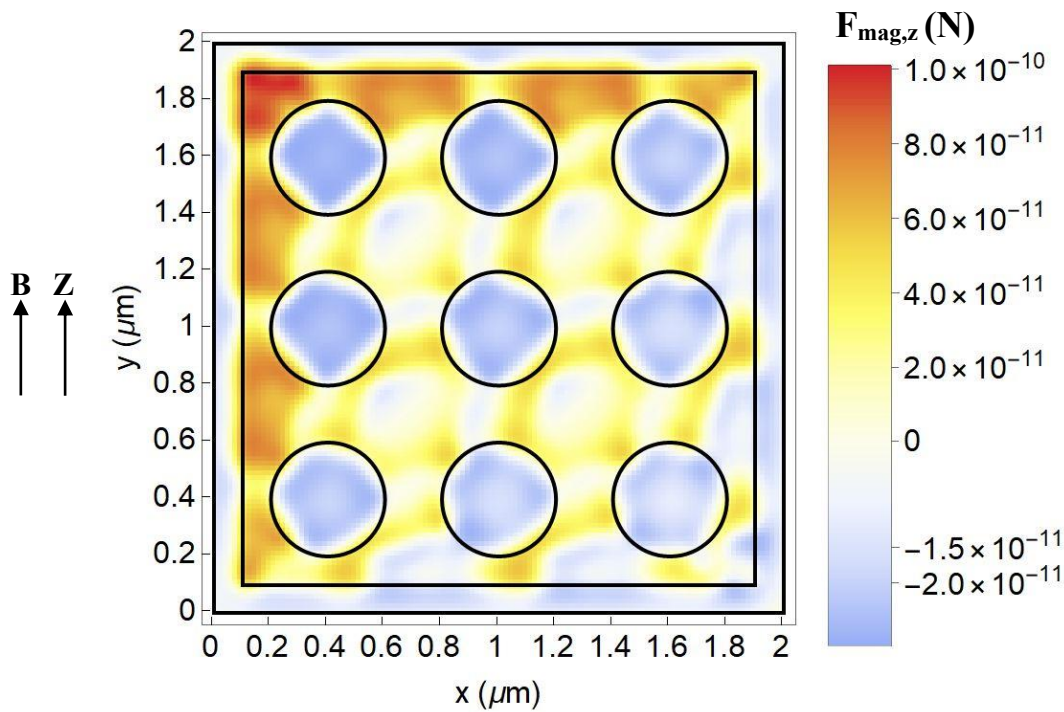

*Figure 9: Two-dimensional plot of the z-component of the magnetophoretic force acting on a 150 nm magnetic nanoparticle, measured in Newtons. The nanoparticle is located 90 nm above the surface of the nanostructure. Circles have also been plotted to indicate where the antidots are located on the nanostructure surface. The external magnetic field B is directed along the z axis with a magnitude of 2 kG. Positive values of the force correspond to MNP attraction to the magnetic nanostructure.*

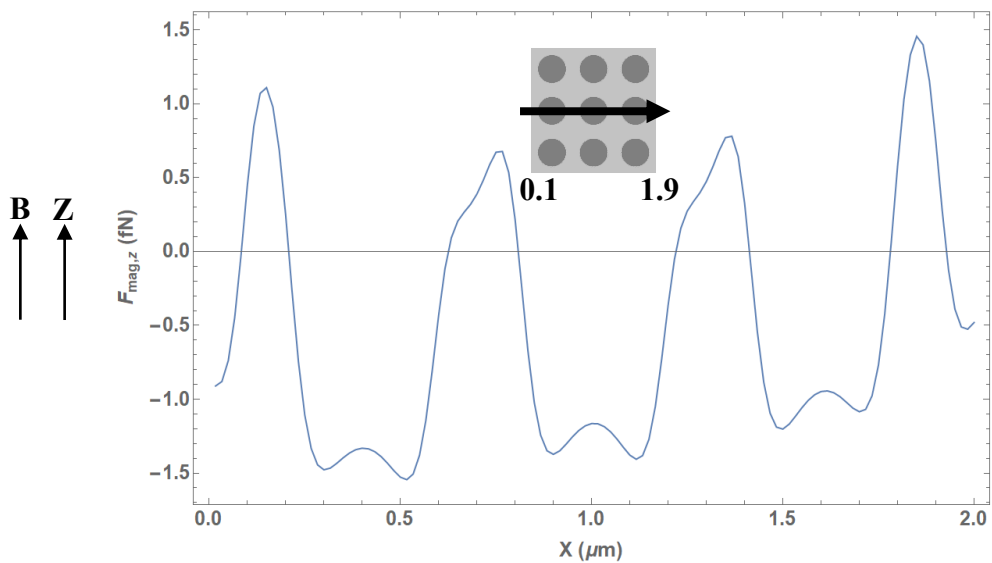

*Figure 10: Simulated z-component of the magnetophoretic force for a 150 nm magnetic nanoparticle located anywhere along the horizontal bisection of the nanostructure shown in the inset. The nanoparticle is located 300 nm above the surface of the nanostructure. The external magnetic field B is directed along the z axis with a magnitude of 2 kG. Negative values of the force correspond to MNP repulsion by the magnetic nanostructure while positive values indicate attraction.*

To determine where the nanoparticles would be captured under an external magnetic field, an estimate of their trajectories is required. The trajectories of a magnetic nanoparticle influenced by the magnetophoretic forces calculated using the above method are shown in Figures 11 - 14 below for both external magnetic fields. A single, motionless MNP was positioned at one of the five locations A to E shown in Figure 11 at a distance of 300 nm from the surface of the nanostructure. The components of the magnetophoretic force acting upon this particle were then calculated and the position of the particle was updated using the Euler method after a time interval of 5 µs. Gravity and hydrodynamic effects such as drag and the flow of the surrounding fluid have been neglected for these simulations.

The trajectories produced under an external magnetic field directed parallel to the surface of the nanostructure are displayed in Figures 11 and 12. The locations B and D between the antidots are regions of repulsive magnetophoretic force according to the force distribution displayed in Figure 6. As anticipated, the particles that start at these positions are repelled away from the nanostructure. Likewise, the locations C and E are regions of attractive magnetophoretic force. The nanoparticles that begin at those positions are attracted towards and into the antidots, passing the outer surface of the nanostructure after 75 to 85 µs. There are no deflections in the x - y plane at location E at the centre of an antidot – here the magnetophoretic force is entirely attractive. Generally, the x and y components of the magnetophoretic force are non-zero and produce motion in the x - y plane. However, the trajectories from positions C and D demonstrate that the z component of the force determines if the nanoparticle is caught by the structure as both travel over an antidot but only trajectory C results in capture.

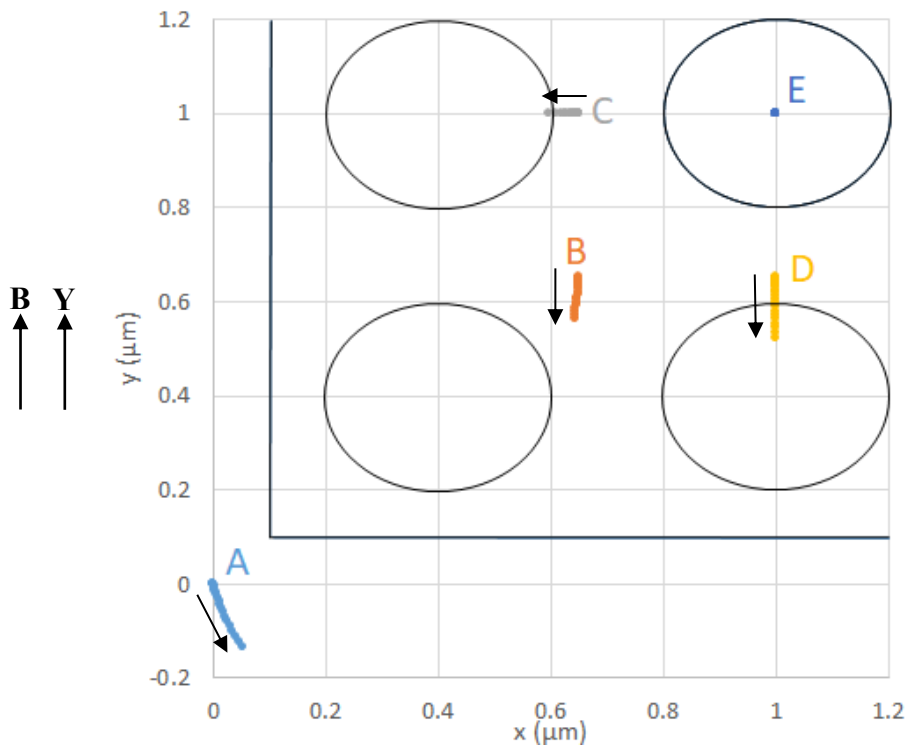

*Figure 11: Trajectories in the xy-plane of a 150 nm magnetic nanoparticle located at positions A to E and starting 300nm above the nanostructure located at z = 0. The external magnetic field B is directed along the y axis with a magnitude of 2 kG.*

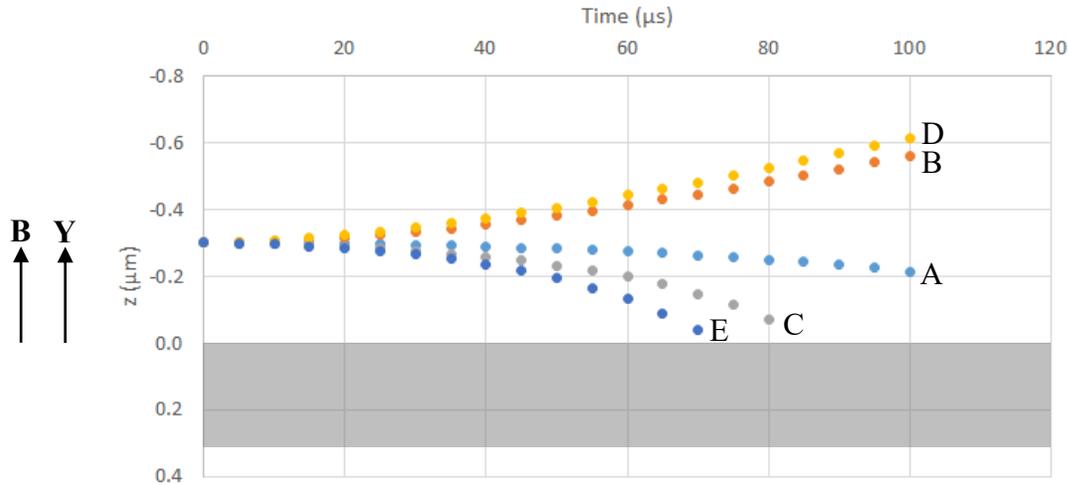

*Figure 12: Trajectories along the z-axis of a 150 nm magnetic nanoparticle located at positions A to E shown in Figure 11 and starting 300nm above the nanostructure. The external magnetic field B is directed along the y axis with a magnitude of 2 kG. The position of the ferromagnetic film has been represented using the gray rectangle. MNPs with negative trajectories move towards the surface of the nanostructure located at z = 300 nm and will be caught by the nanostructure. MNPs with positive trajectories move away from the nanostructure.*

The trajectories produced under an external magnetic field directed perpendicular to the surface of the nanostructure are displayed in Figures 13 and 14. The locations B, C and D between the antidots are all regions of repulsive magnetophoretic force according to the force distribution displayed in Figure 7. Unlike the case with a parallel external magnetic field, the particles that start at these positions are attracted towards the nanostructure and are captured near the centres of their respective regions between antidots. Due to the symmetry of this external field orientation, locations C and D produce identical trajectories along the x and y axes respectively. This is demonstrated in Figure 13 and particularly Figure 14 in which the trajectories for C and D are indistinguishable. As expected, the particle at location E above the antidot is repelled away from the surface. There are no deflections in the x - y plane at this location – here the magnetophoretic force is entirely repulsive.

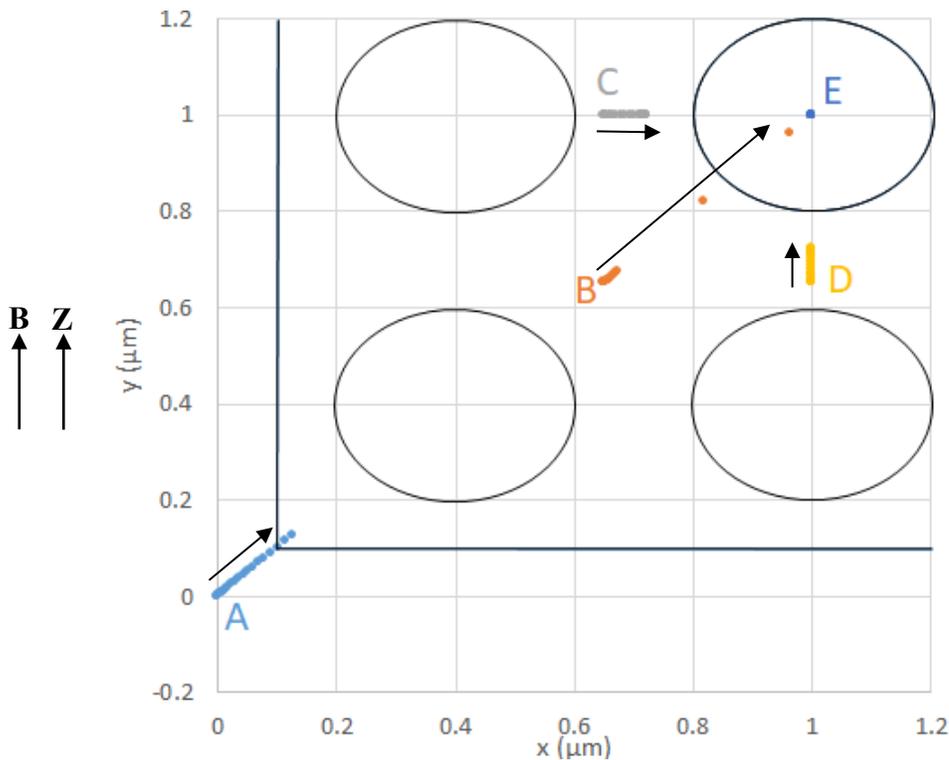

*Figure 13: Trajectories in the xy-plane of a 150 nm magnetic nanoparticle located at positions A to E and starting 300nm above the nanostructure located at z = 0. The external magnetic field B is directed along the z axis with a magnitude of 2 kG.*

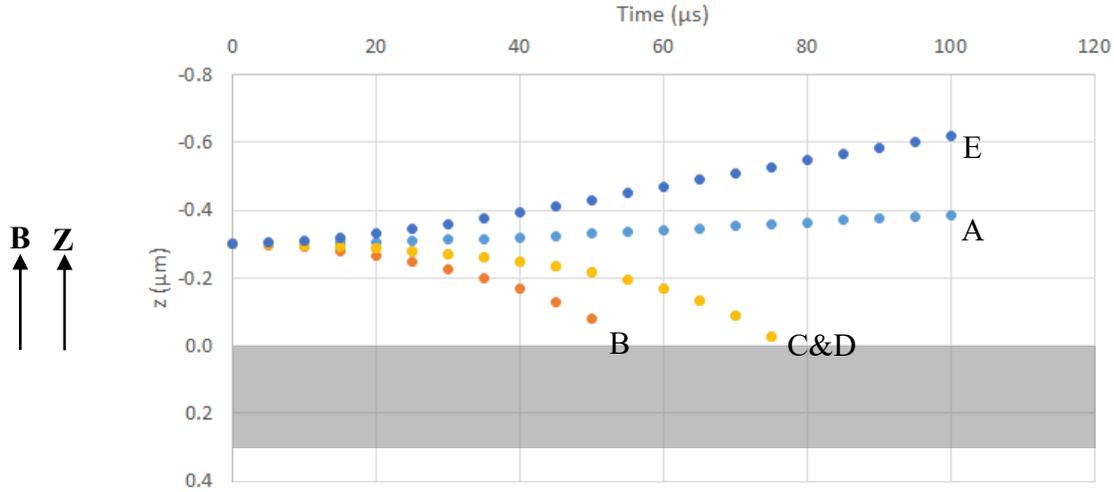

*Figure 14: Trajectories along the z-axis of a 150 nm magnetic nanoparticle located at positions A to E shown in Figure 13 and starting 300nm above the nanostructure. The external magnetic field B is directed along the z axis with a magnitude of 2 kG. The trajectories for positions C and D are identical when only observing motion along the z-axis. The position of the ferromagnetic film has been represented using the gray rectangle. MNPs with negative trajectories move towards the surface of the nanostructure located at z = 300 nm and will be caught by the nanostructure. MNPs with positive trajectories move away from the nanostructure.*

## IV. RESULTS AND DISCUSSION

To verify the results obtained from the simulations, nanostructures of four different geometries were etched into a 30 nm-thick Permalloy film. A droplet of magnetic nanoparticle suspension was then deposited onto the surface of each film using a pipette under an external magnetic field aligned parallel or perpendicular with the films. Once the droplet had dried, SEM images were taken for each of the nanostructures. The SEM images in Figures 15, 16, and 17 display the distributions obtained in one of the antidot array nanostructures without an external field, with a parallel field and with a perpendicular field respectively. Images of MNP distributions in the other nanostructures geometries have been collected into Appendix A. These geometries exhibit qualitatively similar distributions to the distributions of the circular antidot geometry.

### A. Capturing Magnetic Nanoparticles with No External Magnetic Field

When deposited onto the surface of the nanostructure without an external magnetic field, the nanoparticles are distributed relatively evenly across and within the nanostructure. There are fewer nanoparticles in the regions surrounding the patterned area. The local magnetic fields produced by the discontinuous magnetisation result in weak magnetophoretic forces capable of attracting particles towards the structure. The results of the counting for the circular antidots without an external magnetic field are collected in Table I below. The circular antidots had the highest capture efficiency amongst the four geometries at 70.8%. Arrays with dots demonstrated lower capture efficiencies than arrays with antidots despite possessing larger patterned areas.

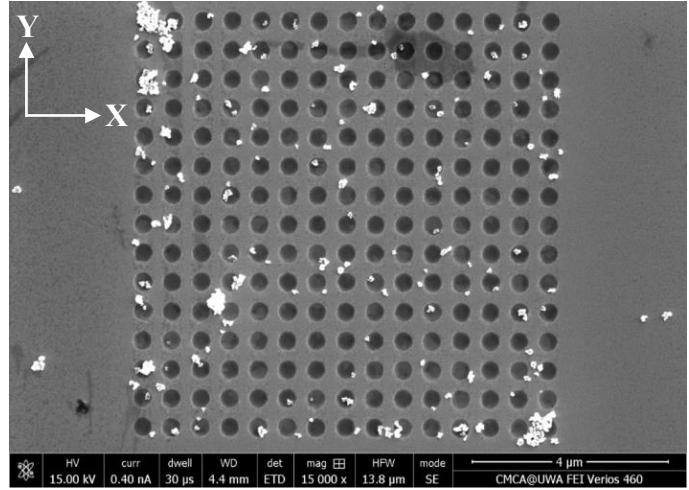

*Figure 15: Distribution of magnetic nanoparticles obtained after a droplet of MNP solution was deposited onto the surface of a circular antidot nanostructure without an external magnetic field.*

TABLE I
MAGNETIC NANOPARTICLES CAUGHT WITHOUT EXTERNAL MAGNETIC FIELD

| Geometry | Particles Inside | Particles Outside | Capture Efficiency (%) |
| --- | --- | --- | --- |
| Circle Antidots | 46.79732 | 19.2796 | 70.82249 |
| Circle Dots | 2.145015 | 7.925353 | 21.30026 |
| Square Antidots | 43.625 | 34.4 | 55.91157 |
| Square Dots | 21.21739 | 38.20652 | 35.70514 |

*Average number of particles caught inside and outside of the nanostructure after deposition without an external magnetic field. Particles trapped 'outside' of the nanostructure are caught on the outer surface of the structure. Conversely, particles trapped 'inside' of the nanostructure are caught on the internal surfaces of the nanostructure. The values shown were weighted by the area of the etched elements relative to the area not etched. The capture efficiency was calculated as the ratio of MNPs caught within the structure to the total number of MNPs caught.*

### B. Capturing Magnetic Nanoparticles with an External Magnetic Field Parallel to the Film

When deposited under an external magnetic field parallel to the surface, the nanoparticles are distributed relatively evenly across the nanostructure. In this external field, fewer particles are caught on the surface of the nanostructure and more particles are caught within the nanostructure. This is in agreement with the results obtained in the simulations of the magnetophoretic forces of these nanostructures. Simulations suggested that the parallel external field would attract more MNPs into the structure, thereby increasing the capture efficiencies. The results of the counting for a parallel external field are presented in Table II. The circular antidot arrays again show the greatest capture efficiency in this field configuration at 70.5%. Circular antidots appear to be the optimal choice for applications requiring MNPs be captured inside of a structure. The efficiencies for the other three geometries are significantly higher in this configuration than without an external field. For example, the lowest efficiency in this configuration was 44.1%, which is more than twice the lowest efficiency of 21.3% when no external field is present. This indicates that the addition of an external magnetic field parallel to the plane of the film substantially improves the capture efficiency of the nanostructure.

Under a parallel field, many nanoparticles began forming larger chain-like clusters. These chains have been observed in other studies using similar magnetic field configurations [38-42]. Self-assembled agglomerations of magnetic nanoparticles are caused by attractive magnetic dipole interactions between particles in close proximity. The size and shape of these chains can be manipulated using the external magnetic field [40]. Smaller, spherical clusters can also be observed in all configurations of the magnetic field and are likely the result of shorter chains collapsing.

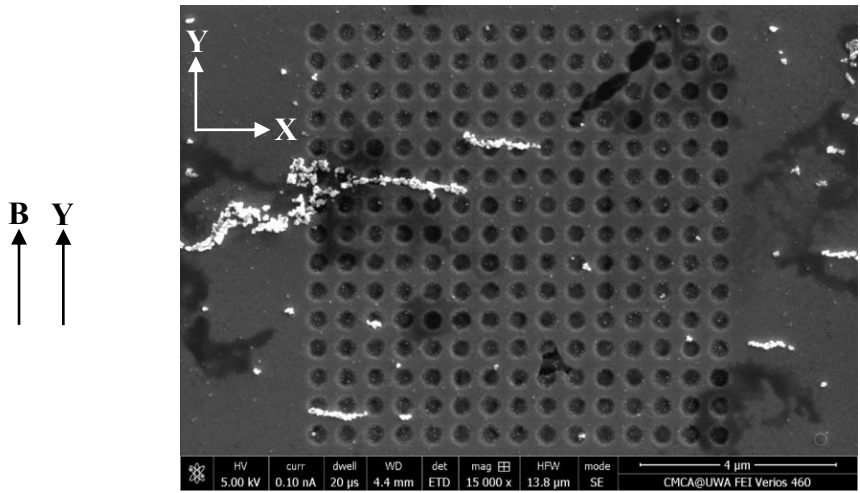

*Figure 16: Distribution of magnetic nanoparticles obtained after a droplet of MNP solution was deposited onto the surface of a circular antidot nanostructure. Deposition occurred within an external magnetic field B directed along the y-axis towards the top of the image.*

TABLE II
MAGNETIC NANOPARTICLES CAUGHT WITH EXTERNAL MAGNETIC FIELD PARALLEL TO FILM

| Geometry | Particles Inside | Particles Outside | Capture Efficiency (%) |
| --- | --- | --- | --- |
| Circle Antidots | 13.30781 | 5.581914 | 70.45 |
| Circle Dots | 6.375461 | 8.080752 | 44.10187 |
| Square Antidots | 13.39286 | 7.580952 | 63.85515 |
| Square Dots | 9.219048 | 10.11905 | 47.67299 |

*Average number of particles caught inside and outside of the nanostructure after deposition with an external magnetic field directed parallel to the surface of the nanostructure. Particles trapped 'outside' of the nanostructure are caught on the outer surface of the structure. Conversely, particles trapped 'inside' of the nanostructure are caught on the internal surfaces of the nanostructure. The values shown were weighted by the area of the etched elements relative to the area not etched. The capture efficiency was calculated as the ratio of MNPs caught within the structure to the total number of MNPs caught.*

### C. Capturing Magnetic Nanoparticles with an External Magnetic Field Perpendicular to the Film

The final magnetic field configuration to consider is an external field directed perpendicular to the surface of the film. In this configuration, the magnetic nanoparticles are no longer distributed evenly across the nanostructure. Most of the particles have accumulated at the edges of the nanostructures and were ignored during all counting. This was predicted by the simulations of the magnetophoretic force due to a perpendicular external field which indicated that the z-component of the force is most attractive at the edges surrounding the nanostructures. In addition, the simulations also predicted that fewer MNPs would be caught within the nanostructure since the magnetophoretic force is repulsive over spaces in the structure. The counting results in Table III confirm this prediction as the capture efficiencies are significantly worse than in the previous external field configurations. Square antidot arrays now exhibit the highest capture efficiency at 28.8% followed by the circle antidots at 24.1%. Antidot arrays still demonstrate slightly better capture efficiencies than the dot arrays. The lowest capture efficiency was only 7.5% for the square dot arrays. This clearly indicates that addition of an external magnetic field perpendicular to the film surface drastically reduces the capture efficiencies of array nanostructures. However, this configuration could be very useful in combining bioseparation and sensing into a single step for µTAS [37, 43]. Additionally, there have been efforts to use magnetic nanoparticles to remove viruses and metal ions for environmentally friendly and cost-effective water treatment [44-46]. In both applications, targets such as viruses could be bound to magnetic nanoparticles and removed from a flowing fluid or isolated for biosensing. By passing the fluid through a number of ferromagnetic sieves, one could catch the viruses bound to particles while allowing everything else in the fluid to pass through.

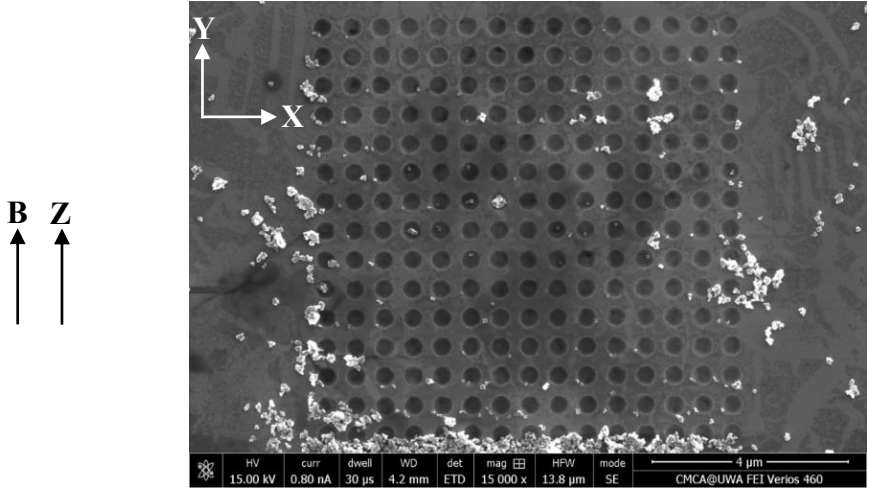

*Figure 17: Distribution of magnetic nanoparticles obtained after a droplet of MNP solution was deposited onto the surface of a circular antidot nanostructure. Deposition occurred within an external magnetic field B directed along the z-axis into the image.*

TABLE III
MAGNETIC NANOPARTICLES CAUGHT WITH EXTERNAL MAGNETIC FIELD PERPENDICULAR TO FILM

| Geometry | Particles Inside | Particles Outside | Capture Efficiency (%) |
|---|---|---|---|
| Circle Antidots | 7.453099 | 23.44891 | 24.11849 |
| Circle Dots | 2.737716 | 33.86064 | 7.480434 |
| Square Antidots | 12.89474 | 31.91579 | 28.77613 |
| Square Dots | 7.621053 | 29.80263 | 20.36425 |

*Average number of particles caught inside and outside of the nanostructure after deposition with an external magnetic field directed perpendicular to the surface of the nanostructure. Particles trapped 'outside' of the nanostructure are caught on the outer surface of the structure. Conversely, particles trapped 'inside' of the nanostructure are caught on the internal surfaces of the nanostructure. The values shown were weighted by the area of the etched elements relative to the area not etched. The capture efficiency was calculated as the ratio of MNPs caught within the structure to the total number of MNPs caught.*

## V. CONCLUSIONS

The results obtained in this study have shown that the capture of magnetic nanoparticles by ferromagnetic nanostructures can be strongly influenced by an external magnetic field and the geometry of the structure. This knowledge may be useful in predicting and manipulating where magnetic nanoparticles are caught. Simulations have shown that applying an external magnetic field parallel to a nanostructure generates magnetophoretic forces that attract the particles into the structure. Conversely, when the external field is directed perpendicular to the structure the nanoparticles are caught outside of the structure. These results were confirmed by depositing magnetic nanoparticles onto the surfaces of nanostructures fabricated using ion beam lithography. In general, circular antidots demonstrated the greatest capture efficiencies. Nanostructures with antidot arrays consistently caught a higher proportion of MNPs than dot arrays. The capture efficiency for four different nanostructure geometries were substantially higher when the external magnetic field was parallel to the film than without an external field. A parallel external field will increase the number of MNPs captured inside of a ferromagnetic nanostructure. It is worth noting that the captured MNPs formed chains in this case. In addition, the nanostructures were found to be significantly less efficient at internally capturing particles when the external field was perpendicular to the structures. In this case, most of the particles were caught on the edges of the structures where the magnetophoretic forces were the most attractive. No chains were formed by the particles that landed away from the edges. Instead, single particles and particle agglomerations in the form of more or less circular lumps were observed. By applying an external magnetic field, one could tailor the capture of magnetic nanoparticles to suit their needs. This may lead to new applications for ferromagnetic nanostructures in the manipulation, imaging and detection of magnetic nanoparticles.

## ACKNOWLEDGEMENTS

R. D. acknowledges a RTP Stipend received from the University of Western Australia. The authors also thank the staff at the Centre for Microscopy, Characterisation and Analysis (CMCA) for access to and assistance with scanning electron microscopy.

SUPPLEMENTARY MATERIAL A – EXPERIMENTAL RESULTS FOR ADDITIONAL NANOSTRUCTURE GEOMETRIES

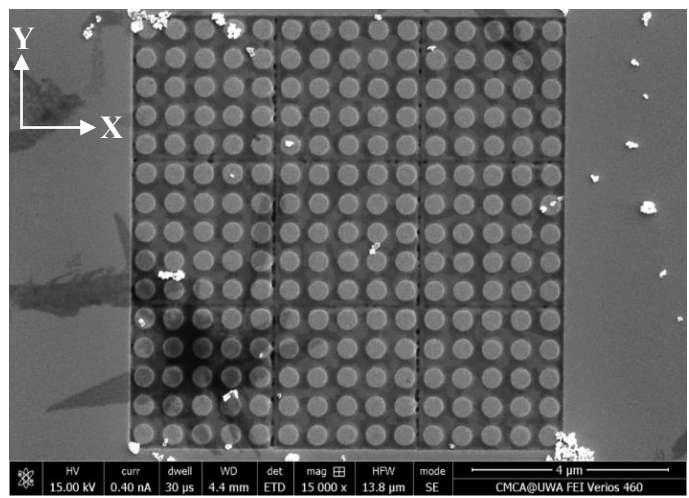

*Figure A1: Distribution of magnetic nanoparticles obtained after a droplet of MNP solution was deposited onto the surface of a circular dot nanostructure without an external magnetic field.*

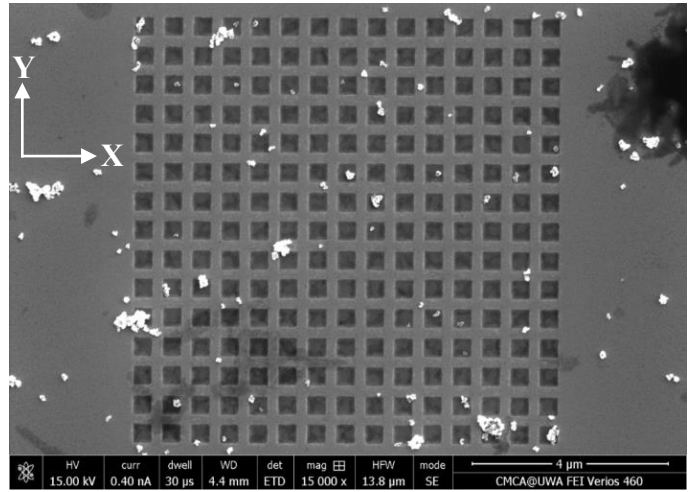

*Figure A2: Distribution of magnetic nanoparticles obtained after a droplet of MNP solution was deposited onto the surface of a square antidot nanostructure without an external magnetic field.*

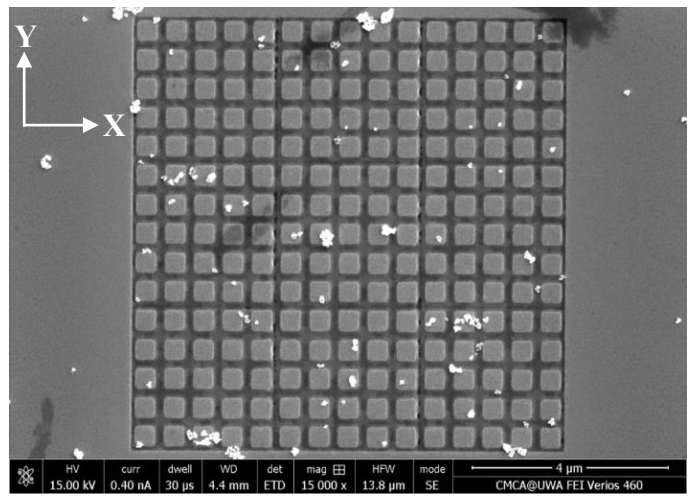

*Figure A3: Distribution of magnetic nanoparticles obtained after a droplet of MNP solution was deposited onto the surface of a square dot nanostructure without an external magnetic field.*

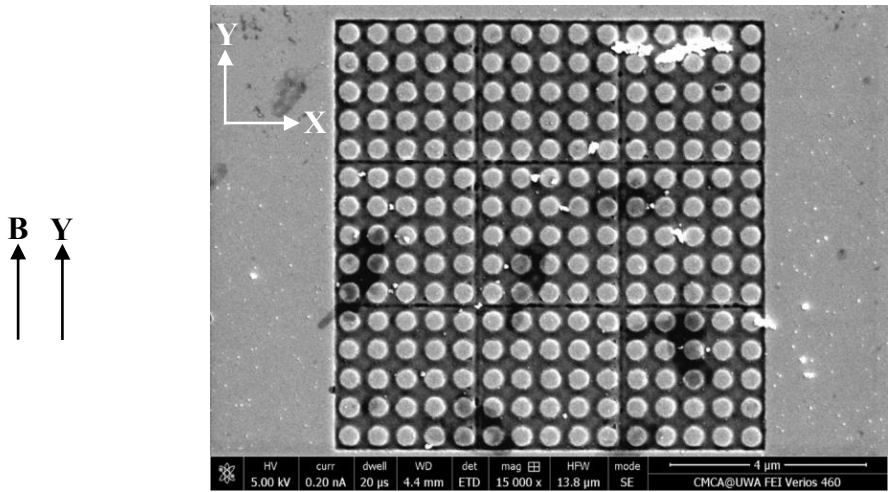

*Figure A4: Distribution of magnetic nanoparticles obtained after a droplet of MNP solution was deposited onto the surface of a circular dot nanostructure. Deposition occurred within an external magnetic field B directed along the y-axis towards the top of the image.*

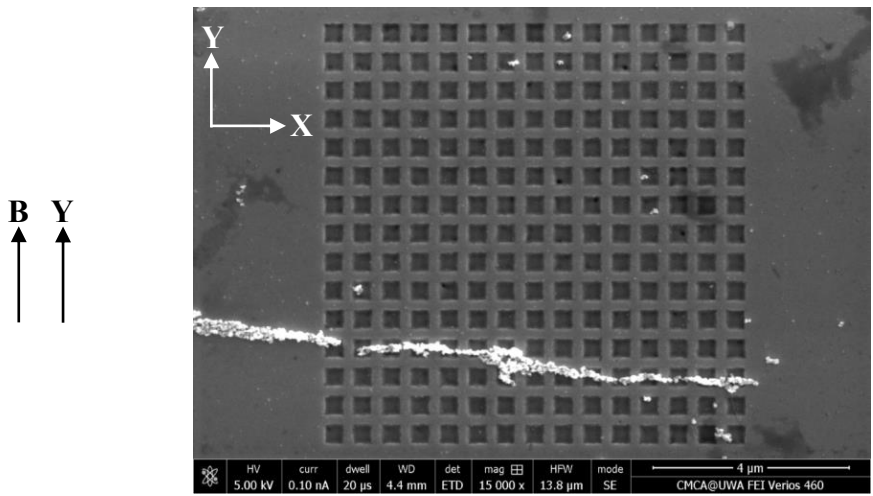

*Figure A5: Distribution of magnetic nanoparticles obtained after a droplet of MNP solution was deposited onto the surface of a square antidot nanostructure. Deposition occurred within an external magnetic field B directed along the y-axis towards the top of the image.*

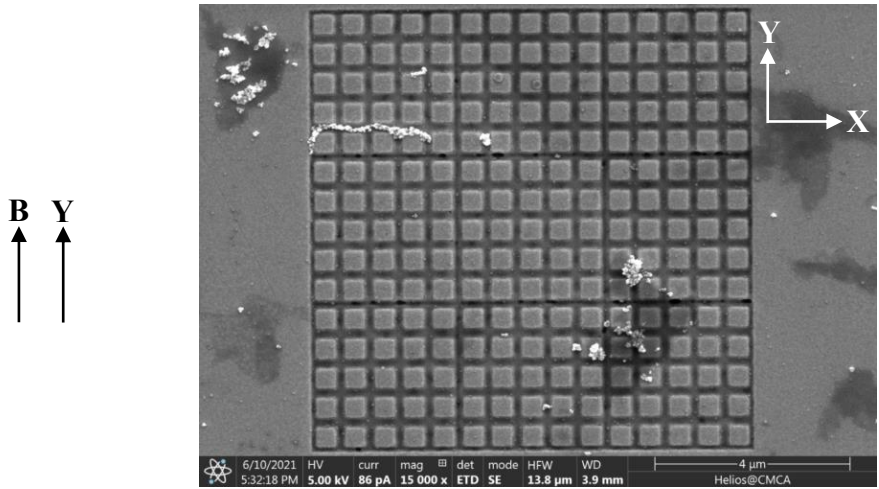

*Figure A6: Distribution of magnetic nanoparticles obtained after a droplet of MNP solution was deposited onto the surface of a square dot nanostructure. Deposition occurred within an external magnetic field B directed along the y-axis towards the top of the image.*

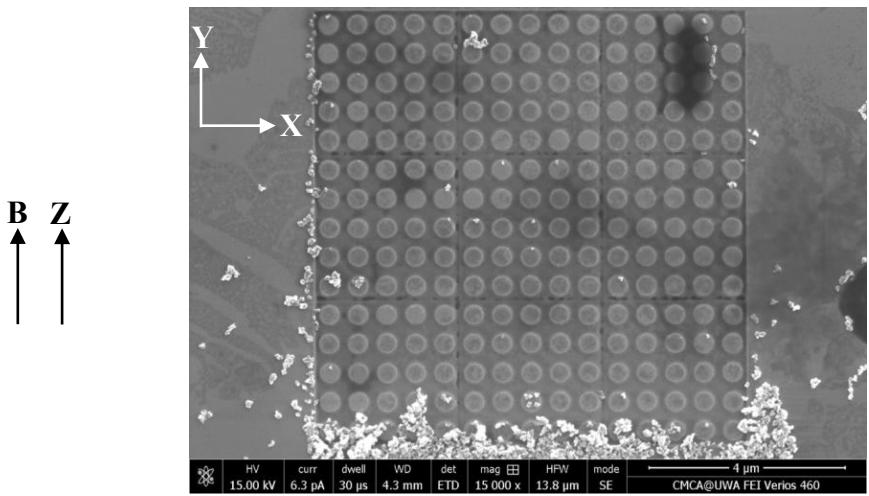

*Figure A7: Distribution of magnetic nanoparticles obtained after a droplet of MNP solution was deposited onto the surface of a circular dot nanostructure. Deposition occurred within an external magnetic field B directed along the z-axis into the image.*

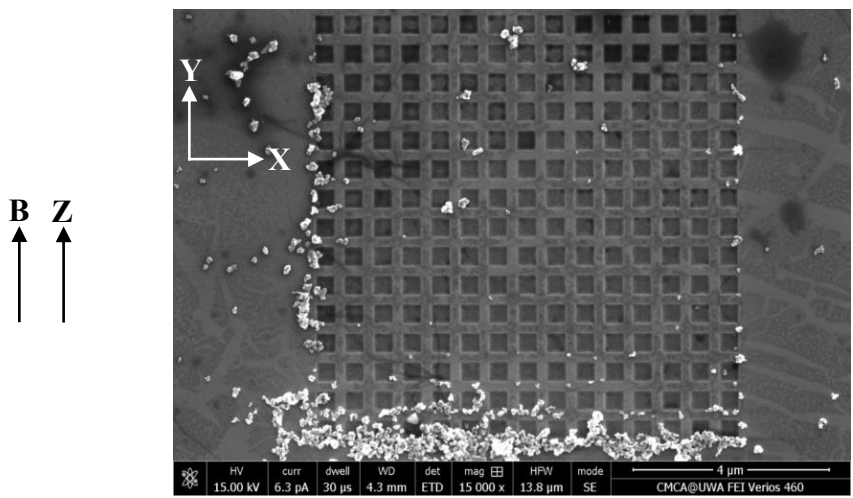

*Figure. A8: Distribution of magnetic nanoparticles obtained after a droplet of MNP solution was deposited onto the surface of a square antidot nanostructure. Deposition occurred within an external magnetic field B directed along the z-axis into the image.*

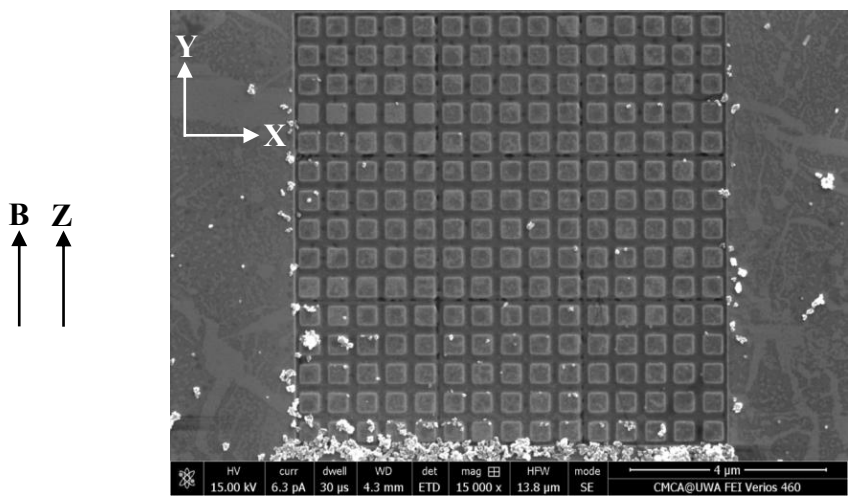

*Figure A9: Distribution of magnetic nanoparticles obtained after a droplet of MNP solution was deposited onto the surface of a square dot nanostructure. Deposition occurred within an external magnetic field B directed along the z-axis into the image.*

SUPPLEMENTARY MATERIAL B – DERIVATION OF ANALYTICAL EQUATIONS FOR THE MAGNETIC FIELD AND MAGNETOPHORETIC FORCE

Suppose there is a nanostructure of arbitrary geometry situated within an external field $\overrightarrow{B_{ex}}$ such that the total field is $\vec{B} = B_x \vec{e_x} + B_y \vec{e_y} + B_z \vec{e_z}$ in Cartesian coordinates. Magnetophoresis is a process in which magnetic nanoparticles are drawn along the gradient lines of this magnetic field $\vec{B}$. The goal is to determine an analytical expression for the magnetophoretic force acting on a magnetic nanoparticle located at the point $P = (x_0, y_0, z_0)$. The magnetophoretic force $\overrightarrow{F_m}$ is given by the equation[25]

$$\overrightarrow{F_m} = A\,\vec{\nabla}|\vec{B}|^2 = A\,\vec{\nabla}(B_x^2 + B_y^2 + B_z^2) = 2\,A\,B_x\vec{\nabla}B_x + 2\,A\,B_y\vec{\nabla}B_y + 2\,A\,B_z\vec{\nabla}B_z \qquad \text{Eq.B1}$$

$$A = \frac{V_{mnp}\,(\chi_{mnp} - \chi_{fluid})}{2\,\mu_0} \qquad \text{Eq.B2}$$

Where $\vec{\nabla}$ is the gradient operator such that $\vec{\nabla}B_x = (\frac{\partial}{\partial x}\vec{e_x} + \frac{\partial}{\partial y}\vec{e_y} + \frac{\partial}{\partial z}\vec{e_z})B_x$. In the constant A, $V_{mnp}$ is the volume of the magnetic nanoparticle, $\mu_0$ is the vacuum permeability and $\chi_{mnp}$ and $\chi_{fluid}$ are the magnetic susceptibilities of the particle and fluid respectively. For these calculations the fluid was chosen to be water with a susceptibility of $9.04 \times 10^{-6}\,m^3$. Eq. B1 indicates that the magnetophoretic force is determined by the magnetic field of the nanostructure. To determine what the total magnetic field will be, consider the material to be comprised of a three dimensional array of magnetic moments located at the points $(x_{ijk}, y_{ijk}, z_{ijk})$ where the indices i, j and k indicate the position in the lattice. Assume that each magnetic moment is a point dipole contributing at point P a magnetic field

$$\overrightarrow{B_{ijk}} = \frac{3\vec{r}(\vec{m}\cdot\vec{r})}{r^5} - \frac{\vec{m}}{r^3} \qquad \text{Eq.B3}$$

Where $\vec{r} = (x_0 - x_{ijk}, y_0 - y_{ijk}, z_0 - z_{ijk}) = (x, y, z)$ is the vector from the dipole to the point of interest P with magnitude $r = \sqrt{x^2 + y^2 + z^2}$ and $\vec{m} = (m_x, m_y, m_z)$ is the magnetisation vector of the point dipole. To get the total magnetic field $\vec{B}$ the external and dipole fields must be summed:

$$\vec{B} = \overrightarrow{B_{ex}} + \sum_{ijk}\vec{B}(x_0 - x_{ijk}, y_0 - y_{ijk}, z_0 - z_{ijk}) = \overrightarrow{B_{ex}} + \sum_{ijk}\overrightarrow{B_{ijk}} \qquad \text{Eq.B4}$$

The total field in Eq. B4 can be substituted into the magnetophoretic force in Eq. B1 to get

$$\overrightarrow{F_m} = 2A(B_{ex,x} + \sum_{ijk}B_{ijk,x})\vec{\nabla}(B_{ex,x} + \sum_{ijk}B_{ijk,x}) + 2A(B_{ex,y} + \sum_{ijk}B_{ijk,y})\vec{\nabla}(B_{ex,y} + \sum_{ijk}B_{ijk,y}) \qquad \text{Eq.B5}$$
$$+ 2A(B_{ex,z} + \sum_{ijk}B_{ijk,z})\vec{\nabla}(B_{ex,z} + \sum_{ijk}B_{ijk,z})$$

In this case the external field is static and the gradient operator is linear so it can be swapped with the summations. The magnetophoretic force therefore simplifies to

$$\overrightarrow{F_m} = 2A(B_{ex,x} + \sum_{ijk}B_{ijk,x})\sum_{ijk}\vec{\nabla}B_{ijk,x} + 2A(B_{ex,y} + \sum_{ijk}B_{ijk,y})\sum_{ijk}\vec{\nabla}B_{ijk,y} \qquad \text{Eq.B6}$$
$$+ 2A(B_{ex,z} + \sum_{ijk}B_{ijk,z})\sum_{ijk}\vec{\nabla}B_{ijk,z}$$

The three components of the magnetophoretic force acting on a particle at point P are

$$\overrightarrow{F_{m,x}} = 2A(B_{ex,x} + \sum_{ijk}B_{ijk,x})\sum_{ijk}\frac{\partial B_{ijk,x}}{\partial x} + 2A(B_{ex,y} + \sum_{ijk}B_{ijk,y})\sum_{ijk}\frac{\partial B_{ijk,y}}{\partial x} \qquad \text{Eq.B7}$$
$$+ 2A(B_{ex,z} + \sum_{ijk}B_{ijk,z})\sum_{ijk}\frac{\partial B_{ijk,z}}{\partial x}$$

$$\overrightarrow{F_{m,y}} = 2A(B_{ex,x} + \sum_{ijk} B_{ijk,x}) \sum_{ijk} \frac{\partial B_{ijk,x}}{\partial y} + 2A(B_{ex,y} + \sum_{ijk} B_{ijk,y}) \sum_{ijk} \frac{\partial B_{ijk,y}}{\partial y} \qquad \text{Eq.B8}$$
$$+ 2A(B_{ex,z} + \sum_{ijk} B_{ijk,z}) \sum_{ijk} \frac{\partial B_{ijk,z}}{\partial y}$$

$$\overrightarrow{F_{m,z}} = 2A(B_{ex,x} + \sum_{ijk} B_{ijk,x}) \sum_{ijk} \frac{\partial B_{ijk,x}}{\partial z} + 2A(B_{ex,y} + \sum_{ijk} B_{ijk,y}) \sum_{ijk} \frac{\partial B_{ijk,y}}{\partial z} \qquad \text{Eq.B9}$$
$$+ 2A(B_{ex,z} + \sum_{ijk} B_{ijk,z}) \sum_{ijk} \frac{\partial B_{ijk,z}}{\partial z}$$

The partial derivatives of Eq. B3 must now be calculated. Introducing the notation

$$m_r = \vec{m} \cdot \vec{r} = m_x x + m_y y + m_z z \qquad \text{Eq.B10}$$

the field produced at point P by each dipole has the components

$$B_{ijk,x} = \frac{3x m_r}{r^5} - \frac{m_x}{r^3} \qquad \text{Eq.B11}$$

$$B_{ijk,y} = \frac{3y m_r}{r^5} - \frac{m_y}{r^3} \qquad \text{Eq.B12}$$

$$B_{ijk,z} = \frac{3z m_r}{r^5} - \frac{m_z}{r^3} \qquad \text{Eq.B13}$$

The first partial derivative to calculate is

$$\frac{\partial B_{ijk,x}}{\partial x} = \frac{3 m_r}{r^5} + \frac{3x}{r^5} \frac{\partial m_r}{\partial x} - \frac{15 x m_r}{r^6} \frac{\partial r}{\partial x} - \frac{1}{r^3} \frac{\partial m_x}{\partial x} + \frac{3 m_x}{r^4} \frac{\partial r}{\partial x} \qquad \text{Eq.B14}$$

Using the definitions for r and $m_r$, we find that $\frac{\partial m_r}{\partial x} = m_x$ and $\frac{\partial r}{\partial x} = \frac{x}{r}$. It is assumed that the external field is strong enough to completely saturate the nanostructure to produce a uniform magnetisation, so $\frac{\partial m_x}{\partial x} = 0$. Substituting these values into Eq. B14 above results in

$$\frac{\partial B_{ijk,x}}{\partial x} = \frac{3 m_r}{r^5} + \frac{3x m_x}{r^5} - \frac{15 x^2 m_r}{r^7} + \frac{3 m_x x}{r^5} \qquad \text{Eq.B15}$$

Which simplifies to

$$\frac{\partial B_{ijk,x}}{\partial x} = \frac{3(m_r + 2x m_x)}{r^5} - \frac{15 x^2 m_r}{r^7} \qquad \text{Eq.B16}$$

$$\frac{\partial B_{ijk,x}}{\partial x} = \frac{3}{r^5}(m_r + 2x m_x - \frac{5 x^2 m_r}{r^2}) \qquad \text{Eq.B17}$$

$$\frac{\partial B_{ijk,x}}{\partial x} = \frac{3}{r^5}(m_r(1 - \frac{5 x^2}{r^2}) + 2x m_x) \qquad \text{Eq.B18}$$

The other partial derivatives can be derived using the same method:

$$\frac{\partial B_{ijk,y}}{\partial y} = \frac{3}{r^5}(m_r(1-\frac{5y^2}{r^2}) + 2ym_y) \quad \text{Eq.B19}$$

$$\frac{\partial B_{ijk,z}}{\partial z} = \frac{3}{r^5}(m_r(1-\frac{5z^2}{r^2}) + 2zm_z) \quad \text{Eq.B20}$$

$$\frac{\partial B_{ijk,x}}{\partial y} = \frac{\partial B_{ijk,y}}{\partial x} = \frac{3}{r^5}(xm_y + ym_x - \frac{5xy}{r^2}m_r) \quad \text{Eq.B21}$$

$$\frac{\partial B_{ijk,x}}{\partial z} = \frac{\partial B_{ijk,z}}{\partial x} = \frac{3}{r^5}(xm_z + zm_x - \frac{5xz}{r^2}m_r) \quad \text{Eq.B22}$$

$$\frac{\partial B_{ijk,y}}{\partial z} = \frac{\partial B_{ijk,z}}{\partial y} = \frac{3}{r^5}(ym_z + zm_y - \frac{5yz}{r^2}m_r) \quad \text{Eq.B23}$$

SUPPLEMENTARY MATERIAL C – NUMERICALLY MODELLED COMPONENTS OF THE MAGNETOPHORETIC FORCE FOR AN ANTIDOT NANOSTRUCTURE

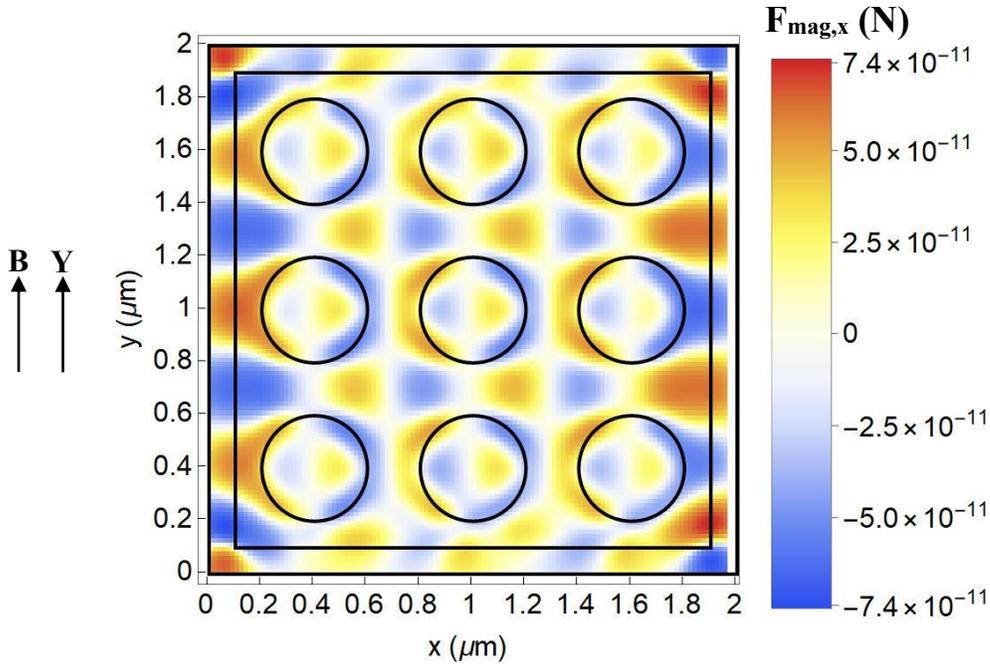

*Figure C1: Two-dimensional plot of the x-component of the magnetophoretic force acting on a 150 nm magnetic nanoparticle, measured in Newtons. The nanoparticle is located 90 nm above the surface of the nanostructure. Circles have also been plotted to indicate where the antidots are located on the nanostructure surface. The external magnetic field B is directed along the y axis with a magnitude of 2 kG.*

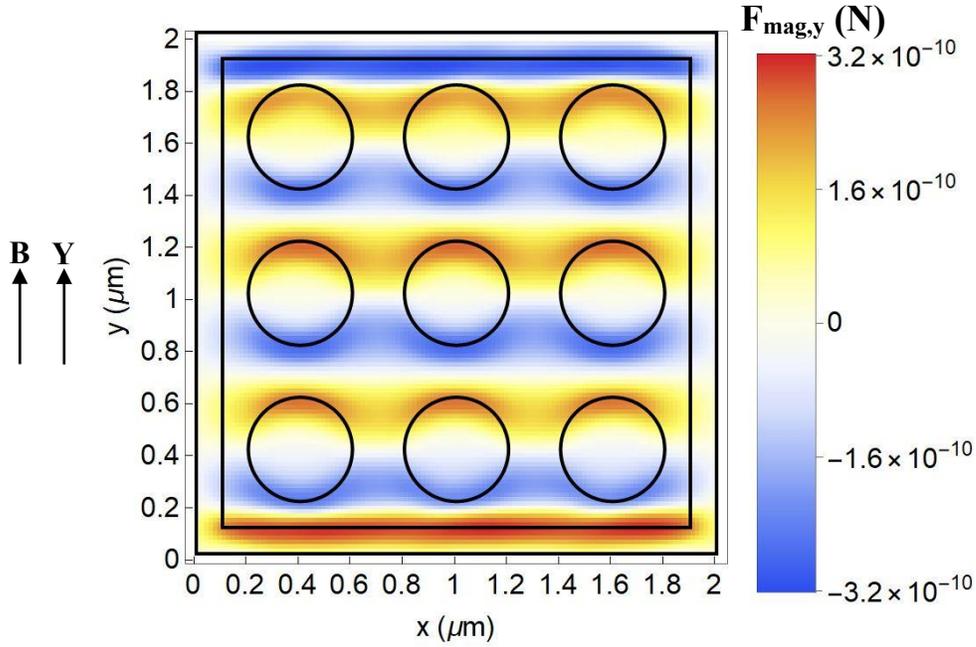

*Figure C2: Two-dimensional plot of the y-component of the magnetophoretic force acting on a 150 nm magnetic nanoparticle, measured in Newtons. The nanoparticle is located 90 nm above the surface of the nanostructure. Circles have also been plotted to indicate where the antidots are located on the nanostructure surface. The external magnetic field B is directed along the y axis with a magnitude of 2 kG.*

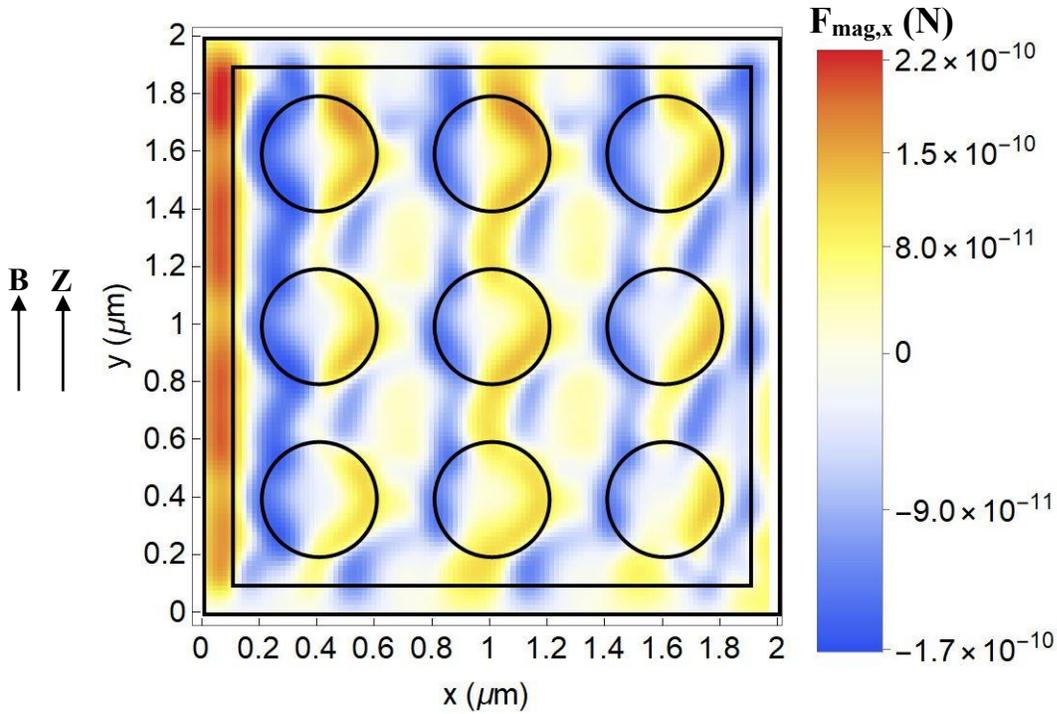

*Figure C3: Two-dimensional plot of the x-component of the magnetophoretic force acting on a 150 nm magnetic nanoparticle, measured in Newtons. The nanoparticle is located 90 nm above the surface of the nanostructure. Circles have also been plotted to indicate where the antidots are located on the nanostructure surface. The external magnetic field B is directed along the z axis with a magnitude of 2 kG.*

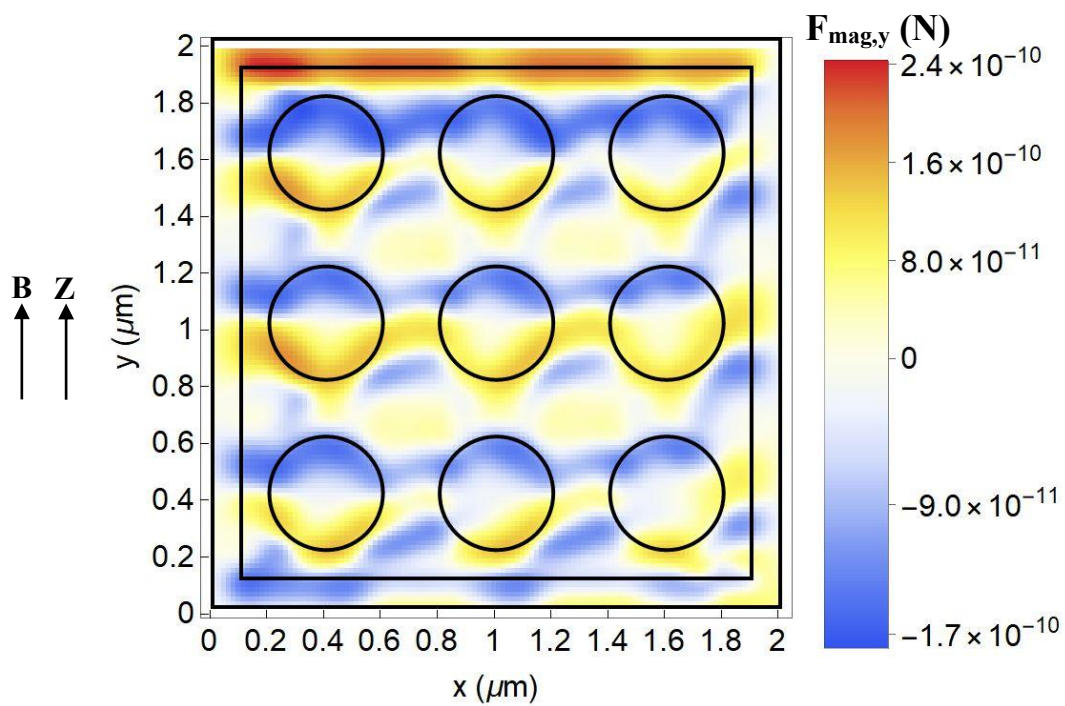

*Figure C4: Two-dimensional plot of the y-component of the magnetophoretic force acting on a 150 nm magnetic nanoparticle, measured in Newtons. The nanoparticle is located 90 nm above the surface of the nanostructure. Circles have also been plotted to indicate where the antidots are located on the nanostructure surface. The external magnetic field B is directed along the z axis with a magnitude of 2 kG.*